\documentclass[A4paper,unsortedaddress,superscriptaddress,prl,showpacs,floatfix,twocolumn]{revtex4}
\usepackage{graphicx}
\usepackage{amsmath}
\usepackage{longtable}
\usepackage[T1]{fontenc}
\newcommand {\TM}[1]{{\bf #1}}

\begin{document}
\DeclareGraphicsExtensions{.pdf}

\title{The molecular structure of the H$_2$O wetting layer on Pt(111)}

\author{Sebastian Standop}
\email{standop@ph2.uni-koeln.de}
\affiliation{II. Physikalisches Institut, Universit\"at zu K\"oln, 50937 K\"oln, Germany} 
\author{Alex Redinger}
\affiliation{II. Physikalisches Institut, Universit\"at zu K\"oln, 50937 K\"oln, Germany}
\altaffiliation[Present address: ]{Universit{\'e} du Luxembourg}
\author{Markus Morgenstern}
\affiliation{II. Physikalisches Institut (IIB) und JARA-FIT, RWTH Aachen, 52056 Aachen, Germany}
\author{Thomas Michely}
\affiliation{II. Physikalisches Institut, Universit\"at zu K\"oln, 50937 K\"oln, Germany}
\author{Carsten Busse}
\affiliation{II. Physikalisches Institut, Universit\"at zu K\"oln, 50937 K\"oln, Germany}

 \date{\today}

\begin{abstract}
The molecular structure of the wetting layer of ice on Pt(111) is resolved using scanning tunneling microscopy (STM). Two structures observed previously by diffraction techniques are imaged for coverages at or close to completion of the wetting layer. At 140\,K only a $\sqrt{37} \times \sqrt{37} \, \text{R25.3} \, ^\circ$ superstructure can be established, while at 130\,K also a $\sqrt{39} \times \sqrt{39} \, \text{R16.1} \, ^\circ$ superstructure with slightly higher molecular density is formed. In the temperature range under concern the superstructures reversibly transform into each other by slight changes in coverage through adsorption or desorption. The superstructures exhibit a complex pattern of molecules in different geometries. 
\end{abstract}


\maketitle

The wetting layer of H$_2$O on metal surfaces is one of the most studied systems in surface science \cite{Feibelman2010}, addressed utilizing different experimental and theoretical techniques on various substrates \cite{Henderson2002,Hodgson2009}. Traditionally, the first adsorbed layer on many hexagonal surfaces of transition metals is rationalized as a single dense-packed ice I$_\text{h}$(0001) plane cut out of the crystal and transferred to the substrate. In a strict interpretation the buckled nature of this plane persists and consequently the ice layer is bound to the metal via an oxygen lone pair of every second molecule whereas one H of every other molecule is pointing upwards (H-up model). However, in the case of Pt(111) there is experimental evidence ruling out that a significant number of free OH groups extend into the vacuum \cite{Ogasawara2002,Haq2002,Petrik2005,Thurmer2008}, leading to a model where the respective H is positioned between O and Pt (H-down model). A third possibility is that the molecular plane is parallel to the surface, i.e. a flat laying molecule. This is also the energetically preferred geometry for monomers on several metals. As a consequence, for Ru(0001) \cite{Haq2006}, Cu(110) \cite{Carrasco2009}, and Pd(111) \cite{Cerda2004}, structures have been observed which maximize the number of such flat lying molecules.

The intuitive structural relation in these systems is a simple commensurate $\sqrt{3} \times \sqrt{3}$ R30$^\circ$ superstructure ($\sqrt{3}$ for short), see \cite{Henderson2002,Hodgson2009} for examples. However, often the strength of the molecule-metal bond is comparable to the molecule-molecule interaction which leads to high-order commensurate (HOC) superstructures, as for H$_2$O/Pt(111): He-Atom-scattering (HAS) \cite{Glebov1997} and low-energy electron diffraction (LEED) \cite{Haq2002,Zimbitas2005} showed a $\sqrt{37} \times \sqrt{37}$ R25.3$^\circ$ superstructure ($\sqrt{37}$ for short) for extended water islands that changes into $\sqrt{39} \times \sqrt{39}$ R16.1$^\circ$ ($\sqrt{39}$ for short) with increasing coverage. The molecular structure, however, remains unsolved. The existing STM data revealed three phases \cite{Morgenstern1997} which could not be reconciled with those found in diffraction \cite{Hodgson2009}. An experimental problem is that the structures are very sensitive and are transformed into the $\sqrt{3}$ by prolonged electron exposure \cite{Haq2002,Harnett2003}. In principle, the structure of the supercell can be calculated using density functional theory (DFT), but for the cell sizes present here, as for an accurate description of hydrogen bonds, this approach is very demanding. Specifically, calculated adsorption energies are often too small to explain wetting \cite{Feibelman2003,Nie2010}. On the basis of molecular resolved images of the pure H$_2$O wetting layer and an analysis of its stability and phase transformations we present here the key elements of the molecular structures of the wetting layer phases.

The experiments were performed in an ultra high vacuum STM system (base pressure $p < 1.5 \times 10^{-11}$~mbar). The sample was prepared by cycles of $5$~keV Ar$^+$ bombardment and flashing ($1270$~K). The azimuthal sample orientation was calibrated using the close-packed steps bounding large vacancy islands created by sputtering at $650$~K. Just before H$_2$O exposure, the sample was flashed to $730 \, \text{K}$ to desorb recaptured adsorbates, especially CO. Thoroughly Ar bubbled, ultrapure H$_2$O was deposited from an Au coated reservoir via a glass tube ending in close proximity to the sample. The dose was calibrated by measuring the pressure in the reservoir of defined volume through a spinning rotor gauge. The typical dose of $0.8$~molecules/Pt site ensured a saturated wetting layer. During and till $300$~s after adsorption the sample was kept at $T_\text{\rm{ads}} \geq 130 \, \text{K}$ which allows adsorbate ordering and re-evaporation of excess molecules not part of the wetting layer \cite{Picolin2009}. To impede subsequent adsorption of impurities, the sample was moved into a cylindrical cold wall (liquid N$_2$) immediately after dosing and cooled to STM imaging temperatures ($20$~K -- $100$~K if not stated otherwise). If these precautions were not obeyed a disordered structure with similar elements of structure resulted. Some of the presented STM images \cite{WSxM} are recorded in differential mode. Due to the inherent drift in STM at variable temperatures, the error in image size is on the order of 5\,\%. Dosing of CO and Xe was performed at $T=20 - 190$~K by backfilling the chamber to $P=1 \times 10^{-9} - 2 \times 10^{-8}$~mbar. The exposure is given in MLE [monolayer equivalent, 1 MLE corresponds to the surface atomic density of Pt(111)].

\begin{figure}[!]
\includegraphics[width=86mm]{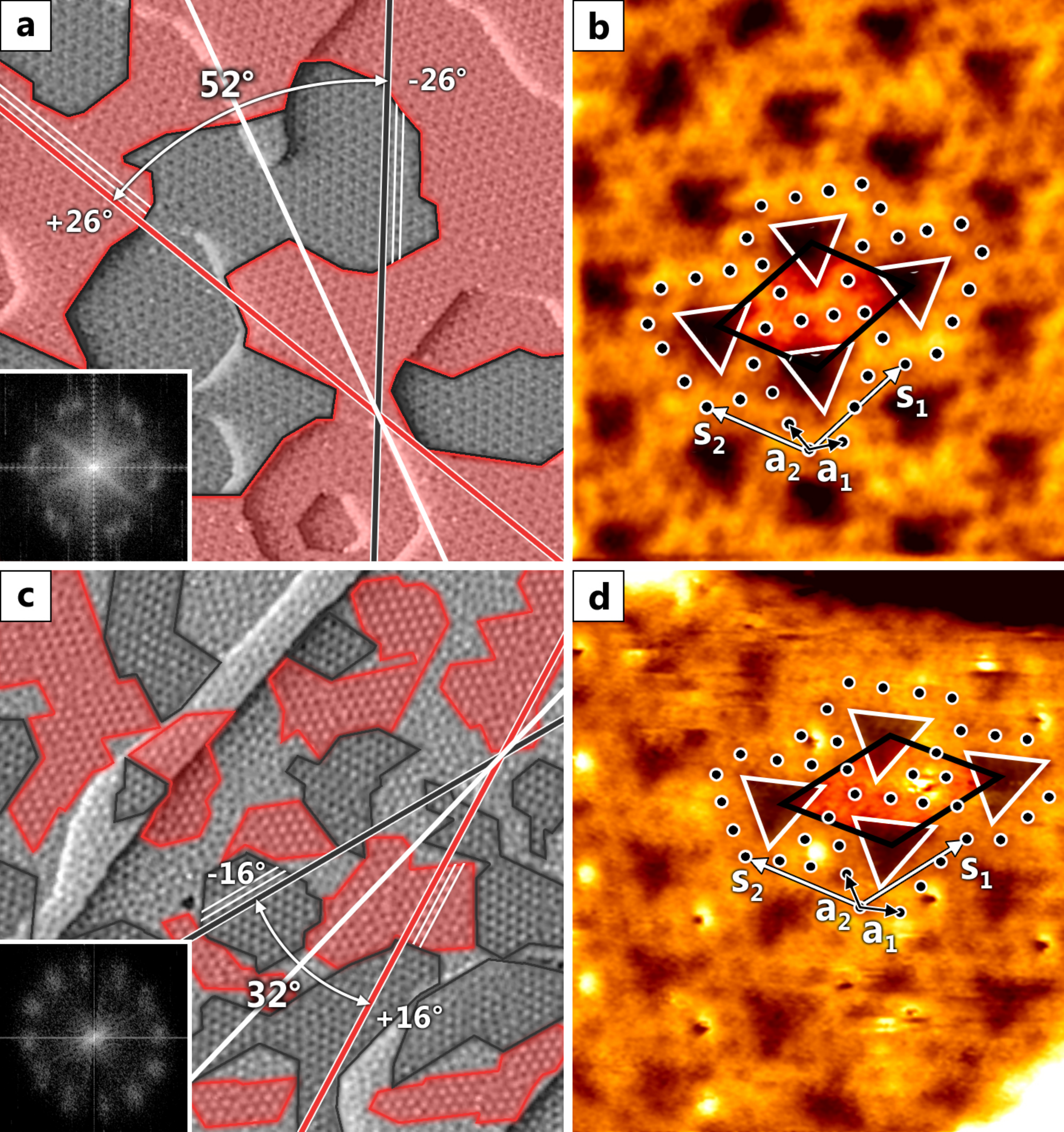}
\caption{(Color online) (a) STM topograph of the H$_2$O wetting layer on Pt(111) formed by adsorption at 140\,K ($\sqrt{37} \times \sqrt{37} \, \text{R25.3} \, ^\circ$). The superstructure is characterized by a lattice of depressions spaced by $17$~\AA. Shading in false colors (dark grey, red) indicates two rotational domains (domain boundaries marked by black and red lines). The lines in the respective colors are parallel to rows of depressions in the appropriate domain (also indicated by parallel thin lines). The angle bisector (white line) is parallel to the step edges of Pt vacancy islands, i.e. parallel to the dense packed $\langle 1\bar{1}0\rangle$-direction. The inset shows the respective Fourier transform. (b) Molecularly resolved STM image of the $\sqrt{37}$ with the supercell marked by a rhombus and associated supercell vectors $\mathbf{s}_1, \mathbf{s}_2$. Triangular depressions are marked by white triangles and lattice points associated with the dark spots are indicated by black dots. Primitive translations of the lattice partly populated by the dark spots are $\mathbf{a}_1$ and $\mathbf{a}_2$. (c) STM topograph of the H$_2$O wetting layer on Pt(111) formed by adsorption at 130\,K ($\sqrt{39} \times \sqrt{39} \, \text{R16.1} \, ^\circ$). The superstructure is characterized by a lattice of depressions and additional protrusions. Color coding, construction of lines, and inset as in (a). (d) Molecularly resolved STM image of the $\sqrt{39}$ showing a periodic second layer decoration (white spots). Markings as in (b). Image parameters: width $710$~\AA (a,c), $81$~\AA (b), $68$~\AA (d), $U = 0.5$~V (a,b,c), $U = 0.2$~V (d), $I = 60$~pA (a), $I = 100$~pA (b), $I = 160$~pA (c), $I=300$~pA (d), width of the Fourier transforms in the insets of (a) and (c) $1.26$~\AA$^{-1}$.}{\label{fig:fig1}}
\end{figure}

Fig.~\ref{fig:fig1}(a) shows the ordered superstructure (appearing as a lattice of depressions) observed after water adsorption at $T_\text{\rm{ads}} = 140 \, \text{K}$ (see \cite{H2O_supplement} for large scale topographs). The Fourier transform displays 12 peaks due to two rotational domains of the hexagonal superstructure. The angle $\alpha = (52 \pm 2)^\circ$ between rotational domains and the periodicity of $(17.1 \pm 0.9) \, \text{{\AA}}$ ($a_{\sqrt{37}} = 16.9 \, \text{\AA}$) indicate that we observe the $\sqrt{37} \times \sqrt{37} \, \text{R25.3} \, ^\circ$ superstructure in real space [equivalent matrix notation $M_{\sqrt{37},{\rm Pt(111)}} = \left( 7 \ 4, -4 \ 3 \right)$ with respect to the substrate, see also Fig.~\ref{fig:fig4}]. In Fig.~\ref{fig:fig1}(b), the molecular structure of the $\sqrt{37}$ is resolved. The triangular depressions are separated by ridges containing dark spots. The dark spots are part of a hexagonal lattice with primitive translations $\mathbf{a}_1, \mathbf{a}_2$. The dense packed rows of this lattice enclose an angle $\beta = 5^{\circ}$ with the ones of Pt(111). With respect to $\mathbf{a}_1$ and $\mathbf{a}_2$ the superstructure has a matrix $M_{\sqrt{37},{\rm H}_2{\rm O}} = \left( 4 \ 2, -2 \ 2 \right)$. We interpret the dark spots as centers of hexagonal ice rings. The structural model proposed in \cite{Glebov1997} does not comply with these observations.

For lower $T_\text{ads} = 130 \, \text{K}$, a second superstructure forms with prominent protrusions [Fig.~\ref{fig:fig1}(b)]. The Fourier transform shows two rotational domains with \TM{$\alpha = (32 \pm 2)^\circ$}. The measured periodicity is $(18.0 \pm 0.5)$~\AA. We identify this structure with the second HOC superstructure: $\sqrt{39} \times \sqrt{39} \, \text{R16.1} \, ^\circ$ ($a_{\sqrt{39}} = 17.3$~\AA) or $M_{\sqrt{39},{\rm Pt(111)}} = \left( 7 \ 5, -5 \ 2 \right)$. Fig.~\ref{fig:fig1}(d) shows that also the $\sqrt{39}$ builds from an ordered arrangement of triangular shaped depressions, but with additional small bright protrusions imaged on the ridges. Here we derive $M_{\sqrt{39}, {\rm H}_2{\rm O}} = \left( 4 \ 3, -3 \ 1 \right)$, with $\beta = 2^{\circ}$. We could also prepare islands of the $\sqrt{37}$ at $130 \, \text{K}$ by adsorbing only 0.55 molecules per site. 

\begin{figure}[!]
\includegraphics[width=86mm]{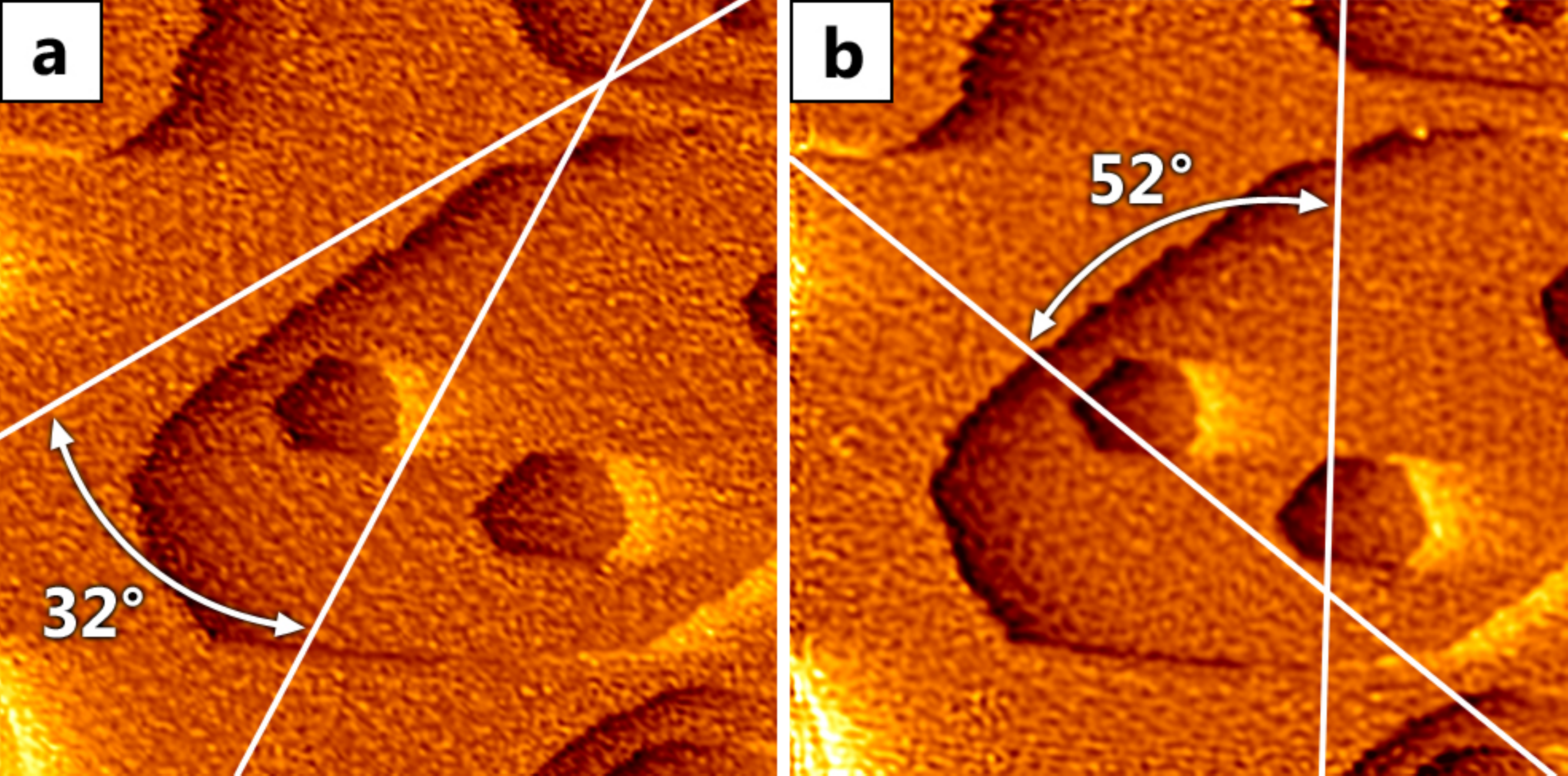}
\caption{(Color online) Phase transition from (a) $T = 138 \, \text{K}$, $\sqrt{39} \times \sqrt{39} \, \text{R16.1} \, ^\circ$ symmetry to (b) $T = 140 \, \text{K}$, $\sqrt{37} \times \sqrt{37} \, \text{R25.3} \, ^\circ$ symmetry within 8 minutes. The white dots vanish, enabling the water layer to reorganize as indicated by the change of the orientation of the superstructure. Image width $820$~\AA, $U = 0.5 \, \text{V}$, $I = 160 \, \text{pA}$.}{\label{fig:fig2}}
\end{figure}

The $\sqrt{37}$ and $\sqrt{39}$ can be reversibly transformed into each other through adsorption or desorption of molecules. One direction of this transformation can be induced by heating the $\sqrt{39}$ to $140$~K. In Fig.~\ref{fig:fig2} the angle between two rotational domains changes from $\alpha = (32 \pm 2)^\circ$ to $\alpha = (52 \pm 2)^\circ$ by heating. Simultaneously, the protrusions of the $\sqrt{39}$ disappear. Together these observations indicate that the $\sqrt{39}$ has a higher density than the $\sqrt{37}$. Once the $\sqrt{37}$ is formed, it is preserved upon cooling. The wetting layer can only be re-transformed from the $\sqrt{37}$ to the $\sqrt{39}$ by subsequent exposure at 130\,K \cite{H2O_supplement}. The reversible transformations indicate that both are equilibrium phases of the wetting layer on Pt(111). The higher thermal stability of the $\sqrt{37}$ implies a higher binding energy per molecule. Moreover, the fact that we never prepared the $\sqrt{39}$ without protrusions shows that they are an integral part of the $\sqrt{39}$.

\begin{figure}[!]
\includegraphics[width=86mm]{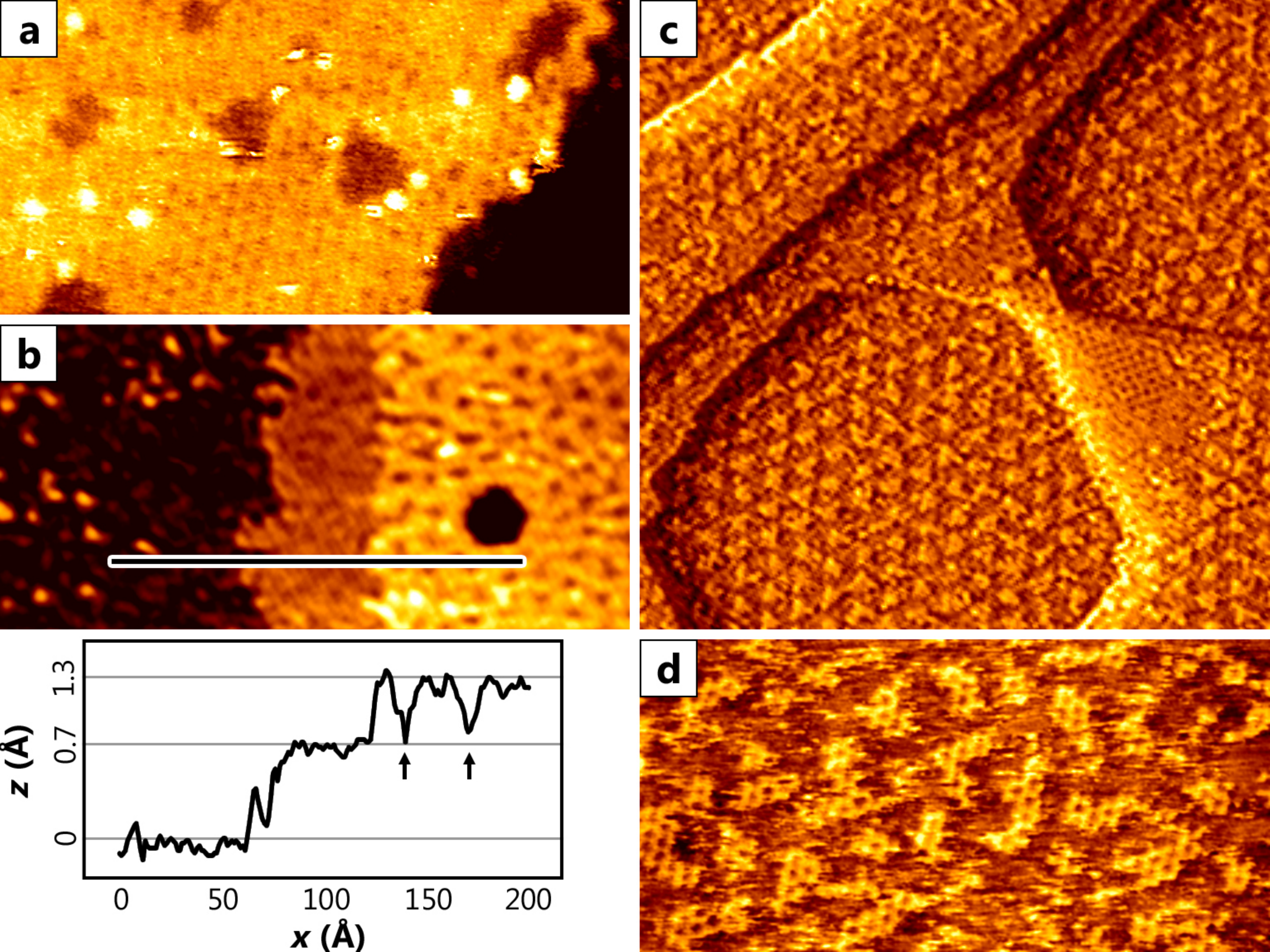}
\caption{(Color online) (a) STM topograph of the water layer after scanning with $U=-5$~V. A $\sqrt{3}$~structure is visible as well as patches of bare Pt. (b) STM topograph next to an area, where the water layer was removed by prolonged scanning at $U=-5$~V. In this transition region the Pt substrate, the dissociated and the intact water layer can be distinguished (see also the respective profile indicated by the black line, black arrows mark the depressions in the water layer). (c) STM topograph in inverted contrast imaging mode (see text) of the water layer after exposure to XXX MLE CO . The triangular depressions are now imaged as triangular protrusions rotated by $30^{\circ}$. (d) With the triangular protrusions as nuclei the $\sqrt{3}$ grows by prolonged scanning in the inverted contrast mode. Image parameters: width $140$~\AA (a), $320$~\AA (b), $310$~\AA (c), $180$~\AA (d), $U = 0.3$~V (a), $U=0.5$~V (b,c,d), $I = 220$~pA (a), $I=190$~pA (b), $I=150$~pA (c,d).}
{\label{fig:fig3}}
\end{figure}

\begin{figure}[!]
\includegraphics[width=86mm]{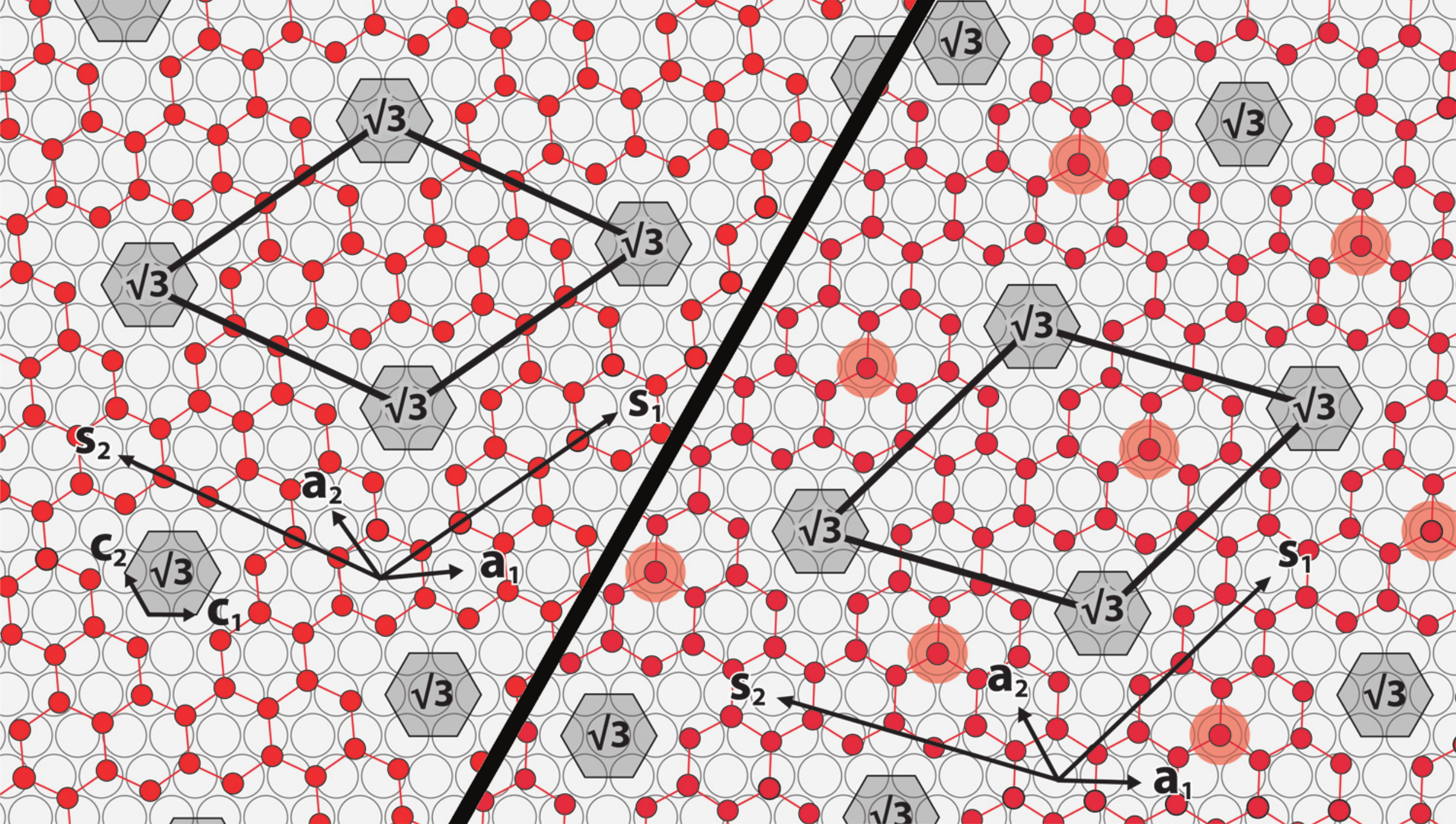}
\caption{(Color online) Sketches of the molecular structure (dismissing hydrogen atoms). Left:  $\sqrt{37} \times \sqrt{37} \, \text{R25.3} \, ^\circ$ supercell. The ridge phase (red circles) is an ice sheet in the H-down geometry, whereas the depressions are segments of a $\sqrt{3}$-structure. The unit cell vectors of the Pt ($\mathbf{c}_1,\mathbf{c}_2$), the H$_2$O lattice ($\mathbf{a}_1,\mathbf{a}_2$) and the superstructure ($\mathbf{s}_1, \mathbf{s}_2$) are indicated. Right: Model of the $\sqrt{39} \times \sqrt{39} \, \text{R16.1} \, ^\circ$ supercell. Symbols as in a), in regions where the 2nd layer admolecules are found are marked by light red circles.}{\label{fig:fig4}}
\end{figure}

The wetting layer can be modified by prolonged scanning with $|U| \geq 4$~V as shown in Fig.~\ref{fig:fig3} (a). First a $\sqrt{3}$ structure results, as typical for H$_2$O/Pt(111) after electron irradiation, usually explained by electron-induced partial dissociation of H$_2$O \cite{Haq2002,Harnett2003}. STM-induced dissociation is also known from related systems \cite{Mehlhorn2008}. Continued scanning leads to the complete removal of the water layer. At the edge of water-free areas [Fig.~\ref{fig:fig3} (b)] a transition from the Pt surface (left) via the dissociated $\sqrt{3}$ (middle) to the $\sqrt{37}$ (right) can be observed. Three associated height levels can be determined in the respective line profile: For $U=0.5$~V, the dissociated phase has a height of $(0.7 \pm 0.1)$~\AA, whereas the $\sqrt{37}$ has a height of $(1.4 \pm 0.1)$~\AA. The apparent height of the $\sqrt{37}$ varies smoothly with tunneling voltage (from $1.3-2.4$~\AA). For a clean tip it shows no contrast inversion, comp. \cite{Morgenstern2004}. The depressions in the intact layer have an apparent depth between $0.6$~\AA and $1.0$~\AA with a mean value of $(0.7 \pm 0.1)$~\AA, the error mainly given by the varying sharpness of the tip. Note that the bottom of the triangular depressions has approximately the same height as the $\sqrt{3}$-structure.

In the following we analyze the nature of the depressions as the common structural element. Several examples of an inhomogeneous appearance of an H$_2$O wetting layer have been reported. In \cite{Morgenstern1997,Thurmer2008} the structure was interpreted as a moir{\'e}-pattern, i.e. an interference pattern arising when two similar grids are superimposed. However, this effect can be ruled out here as it cannot explain the different appearance of neighboring depressions, the rather large depth, or the sharp edges of these pits. Another possible explanation is that the depressions are molecular vacancy clusters, similar to the lace structure reported for sub-monolayer coverages of H$_2$O on Pd(111) \cite{Cerda2004}, which indeed bears a striking resemblance to the structures observed here. Geometrically, in such a flat hexagonal structure an imbalance between hydrogen bond donors and acceptors arises which leads to the appearance of holes in the water film.

About 5\,\% of the saturation coverage of CO adsorbs to the $\sqrt{39}$ and only a two layer thick ice film has a zero sticking probability for CO \cite{Harnett2003}. Does this observation corroborate the hole-model? We tested this by dosing CO on a closed water layer. However, the appearance of the triangular depressions did not change even upon exposure in excess of 1 MLE CO. Instead, widely spaced islands of the CO $\sqrt{3} \times \sqrt{3}$- or c$(4 \times 2)$-structure \cite{Ertl1977} appeared inside the water layer [visible in Fig.~\ref{fig:fig3} (c) right from the center) indicating that CO is able to partly break up the H$_2$O layer here. Also for Xe no change in the appearance of the triangular depressions was observed.

During the CO-experiments occasionally an inverted contrast STM imaging mode was encountered. We tentatively attribute this imaging mode to a CO molecule adsorbed to the PtIr tip. In inverted contrast mode the triangular depressions are imaged as protrusions with their triangular envelope rotated by $30^{\circ}$. The other parts of the wetting layer and CO-islands are imaged $0.8$~\AA~lower than the protrusions. Prolonged scanning in the inverted contrast mode causes the triangular protrusions in the wetting layer to grow to a connected pattern of the $\sqrt{3}$ [Fig.~\ref{fig:fig3} (d)], which displays the same height as the original protrusions. Apparently, the triangular protrusions are nuclei of the $\sqrt{3}$.

Based on the experimental results we suggest the following elements for the $\sqrt{37}$ and $\sqrt{39}$ as shown in Fig.~\ref{fig:fig4}. (i) The ridges of the superstructures consist of a hexagonal network of ice molecules derived from the lattice generated by the primitive translations $\mathbf{a}_1$ and $\mathbf{a}_2$ in Fig.~\ref{fig:fig1}. Each lattice point visible as a dark spot in Figs.~\ref{fig:fig1}(b) and (d) is surrounded by an ice ring. (ii) The triangular depressions are filled with water molecules forming the basic building block of the $\sqrt{3}$ or something very similar (gray hexagons in Fig.~\ref{fig:fig4}). (iii) In addition, to the $\sqrt{39}$ in each unit cell a second layer water molecule is adsorbed.

(i) The proposed arrangement of the H$_2$O lattice {ridges} with respect to the substrate is backed by the evaluation of the spot intensities found in previous diffraction studies \cite{Glebov1997,Haq2002,Zimbitas2005}: The spots with the highest intensities coincide with the reciprocal lattice points calculated from the partly populated black dot lattice in our model. Note that this is not the case for the structure proposed originally \cite{Glebov1997} (for more details see \cite{H2O_supplement}). Consistent with previous results the ridges of the intact wetting layer are assumed to consist of a structure according to the H-down model \cite{Ogasawara2002,Haq2002,Petrik2005,Thurmer2008}. Previous work rules out that a significant number of dangling hydrogen bonds points into the vacuum, which makes the first water layer hydrophobic, exemplified by 3D ice growth on top. Consequently, we propose that the molecules in the ridges are arranged similar to the H-down model. However, as the position of the hydrogen atoms cannot be determined using STM we made no attempt to propose an orientation of individual water molecules shown in Fig.~\ref{fig:fig4}. Furthermore, we made no attempt to relax the molecular positions. In the resulting $\sqrt{37}$ ($\sqrt{39}$), the hydrogen bond length of the molecules in the ridges is extended only by $0.7 \, \%$ ($1.8 \, \%$) compared to the one in bulk ice even if one neglects possible relaxations through the presence of the depressions. (ii) The equilibrium structure of a 1:1 mixed $(\sqrt{3} \times \sqrt{3})$~R$30^{\circ}$ OH + H$_2$O structure is composed out of OH and H$_2$O in the same height above the substrate and with their molecular planes oriented parallel to the surface, i.e. a flat-lying geometry \cite{Michaelides2001b}. As we find the same adsorption geometry and a comparably low apparent height after dissociation we interpret the structure observed in Fig.~\ref{fig:fig3} (a) as this flat-lying phase. Adsorption of H$_2$O under impure conditions also leads to $\sqrt{3}$ regions within the water layer which is of similar height than the wetting layer and thus distinct from the flat-laying $\sqrt{3}$ regions. As the triangular depressions are not empty, have under regular imaging conditions the same height as the $\sqrt{3}$ and act as their nuclei they are most probably filled with flat-lying molecules and/or flat-lying fragments in an $\sqrt{3}$-geometry. It has to be noted that a small flat-lying segment does not have to contain dissociated H$_2$O in order to avoid the occurrence of frustrated hydrogen bonds. If a complete hexagon of a non-dissociated $\sqrt{3}$-phase is placed inside the depression one arrives at the structures recently proposed in \cite{Nie2010}. (iii) For the $\sqrt{39}$ an additional molecule per unit cell can adsorb on top of the wetting layer by saturating dangling bonds, and thus lowering the energy \cite{Zimbitas2008}. This provides a simple explanation for the bright protrusions observed in Fig.~\ref{fig:fig1} (d).

In order to obtain a coverage estimate, we assume the same molecular density in the triangular depressions as in the ridge phase, leading to $0.65$ ($0.70$) molecules/site  for $\sqrt{37}$ ($\sqrt{39}$), in fair agreement with uptake measurements \cite{Clay2004b} for the $\sqrt{39}$. The ease of the phase transformation can be rationalized by the similarity of the two structures, the almost identical length of the OH-bonds and the similar orientation of the dense packed molecular rows.  Note that the $\sqrt{37}$ is most probably identical to phase II in \cite{Morgenstern1997}, whereas phase I from this reference could not be reproduced.

\begin{acknowledgments}
The authors acknowledge critical reading of the manuscript and technical advice by G. Pirug as well as discussions with P. J. Feibelman and the receipt of a DFT calculation for the $\sqrt{37}$ structure in December 2009 on the basis of  our structure, but with empty triangular depressions. Also receipt of \cite{Nie2010} on April 13, 2010 is acknowledged. This study was financially supported by Deutsche Forschungsgemeinschaft.
\end{acknowledgments}

\end{document}


\DeclareGraphicsExtensions{.pdf}

\title{The molecular structure of the H$_2$O wetting layer on Pt(111) - Supplementary material}

\author{Sebastian Standop}
\email{standop@ph2.uni-koeln.de}
\affiliation{II. Physikalisches Institut, Universit\"at zu K\"oln, 50937 K\"oln, Germany} 
\author{Alex Redinger}
\affiliation{II. Physikalisches Institut, Universit\"at zu K\"oln, 50937 K\"oln, Germany}
\altaffiliation[Present address: ]{Universit{\'e} du Luxembourg}
\author{Markus Morgenstern}
\affiliation{II. Physikalisches Institut (IIB) und JARA-FIT, RWTH Aachen, 52056 Aachen, Germany}
\author{Thomas Michely}
\affiliation{II. Physikalisches Institut, Universit\"at zu K\"oln, 50937 K\"oln, Germany}
\author{Carsten Busse}
\affiliation{II. Physikalisches Institut, Universit\"at zu K\"oln, 50937 K\"oln, Germany}

 \date{\today}

 \maketitle
 
\begin{figure}[htb]
\includegraphics[width=\textwidth]{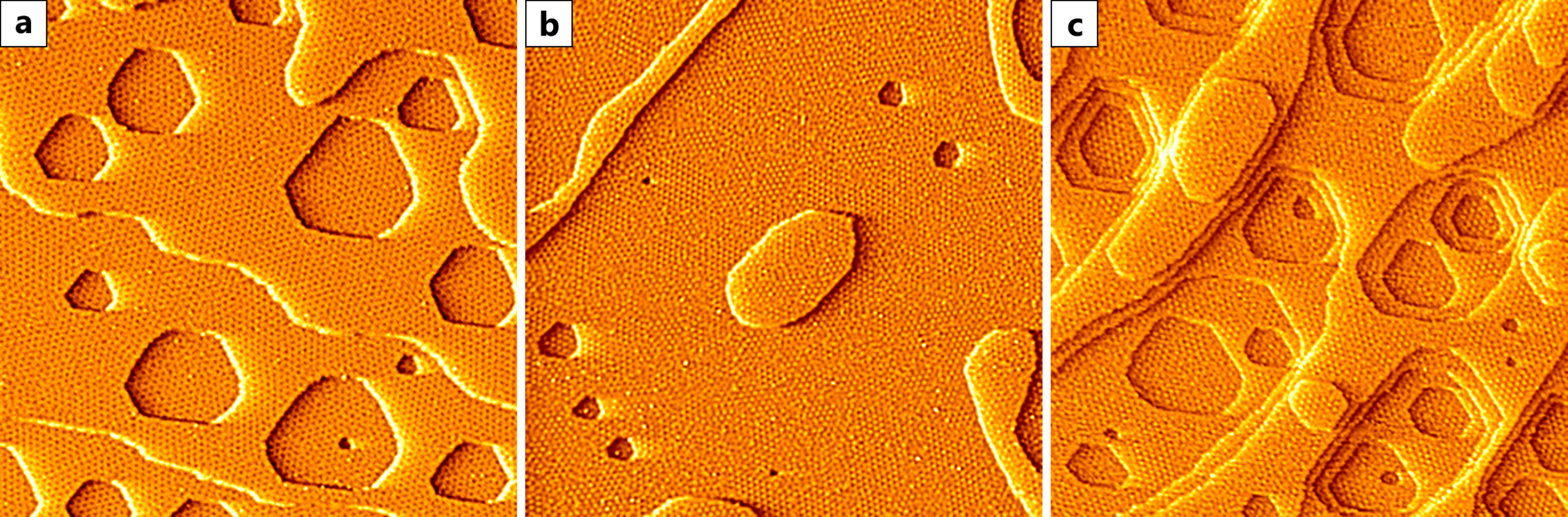}
\caption{(a) STM topograph of the H$_2$O wetting layer on Pt(111) formed by adsorption at 140\,K and (b) 130\,K. The superstructures are characterized by a lattice of (a) depressions ($\sqrt{37} \times \sqrt{37} \, \text{R25.3} \, ^\circ$) and (b) additional protrusions ($\sqrt{39} \times \sqrt{39} \, \text{R16.1} \, ^\circ$). (c) $\sqrt{39} \times \sqrt{39} \, \text{R16.1} \, ^\circ$ formed obtained via a phase transformation starting from a $\sqrt{37} \times \sqrt{37} \, \text{R25.3} \, ^\circ$  formed by deposition at $140$~K as in (a). The phase transformation was initiated by subsequent deposition at $130$~K. The image shows a different spot on the sample as compared with (a). Image parameters: Width $1200$~\AA, $U=1,0$~V (a), $U=0,5$~V (b,c), $I=160$~pA (a), $I=160$~pA (b), $I=150$~pA (c).{\label{fig:largescale}}}
\end{figure}

The long range order obtained in our experiments is shown in Fig.~\ref{fig:largescale}. At $T_\text{ads} = 140 \, \text{K}$ [Fig.~\ref{fig:largescale}(a)] the superstructure domains appear as hexagonal patterns of depressions. At $T_\text{ads} = 130 \, \text{K}$ [Fig.~\ref{fig:largescale}(b)] the superstructure domains show additional protrusions, imaged as bright spots decorating the superstructure unit cell.


Complementary to the phase transformation from $\sqrt{39}$ to $\sqrt{37}$ induced by heating the sample (Fig.~2 of the main paper) also the reverse process can be observed. However, for this phase transition (from a $\sqrt{37}$ to a $\sqrt{39}$ wetting layer) it is not sufficient to cool the sample back towards $T_\text{s} = 130 \, \text{K}$. Once the $\sqrt{37}$ layer is formed, it is preserved even at lowest temperatures. As discussed in the main text, the phase transition upon sample heating is accompanied by desorption of individual molecules. Cooling the sample without offering additional water molecules does not account for reversible processing. To re-transform the wetting layer from $\sqrt{37}$ to $\sqrt{39}$ cooling has to be followed by water deposition. The resulting images are presented in Fig.~\ref{fig:largescale}. We can conclude, that these high order commensurate superstructures are equilibrium phases.

\begin{figure}[htb]
\includegraphics[width=\textwidth]{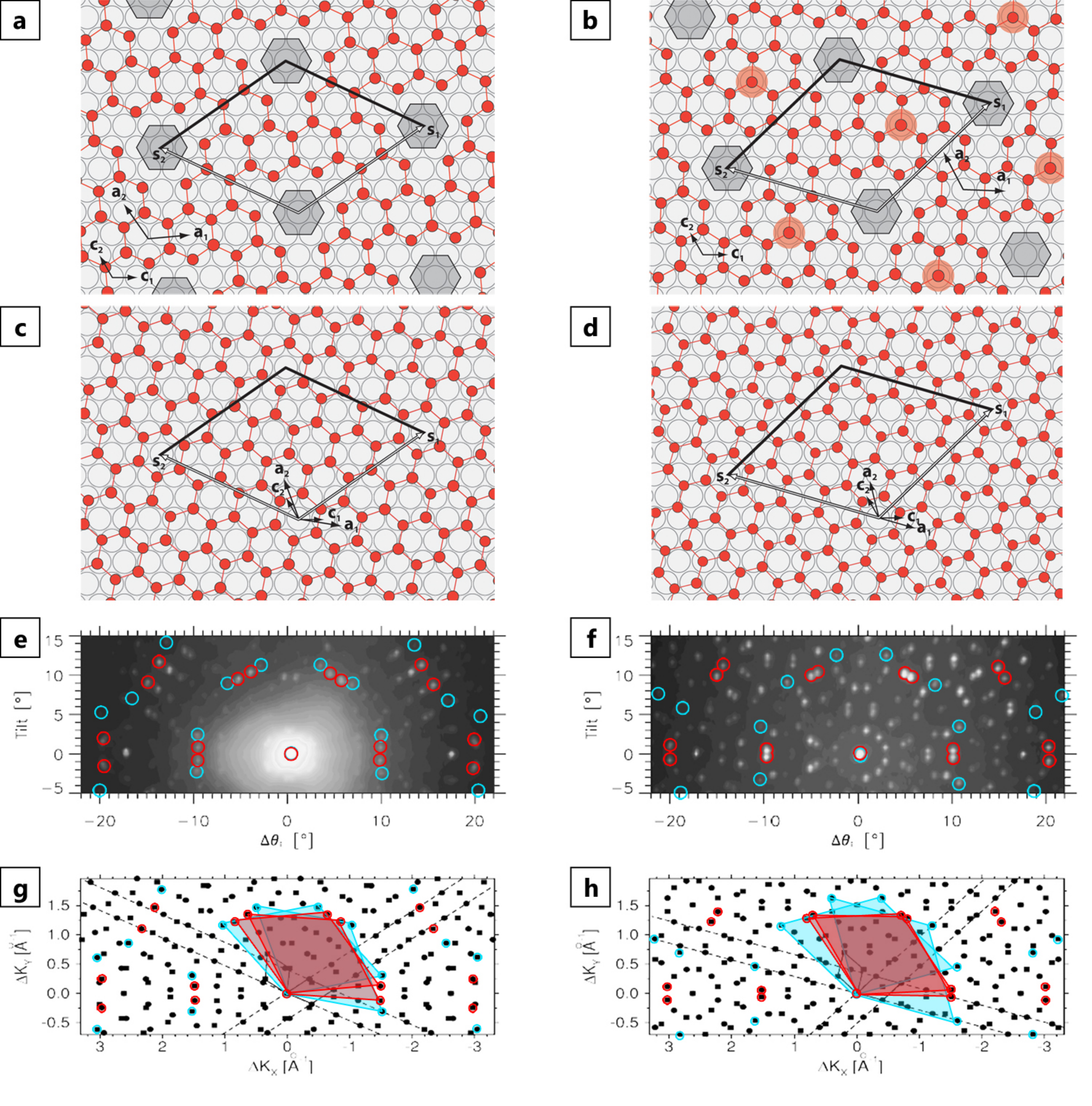}
\caption{(a),(b) Structures of the water layer on Pt(111) proposed in this publication, (a) $\sqrt{37}$, (b) $\sqrt{39}$. See text of main paper for more information. (c), (d) Sketch of the molecular arrangement suggested in \cite{Glebov1997} on basis of HAS data, (c) $\sqrt{37}$, (d) $\sqrt{39}$. (e), (f), HAS data originally published in \cite{Glebov1997} overlayed with colored circles (blue: Glebov-model, red: structure proposed by us) indicating where reflexes that are expected to be of high intensity according to our simplified approach should be found (see text). (e) $\sqrt{37}$, (f) $\sqrt{39}$. (g), (h) Schematic construction of the diffraction pattern as originally published in \cite{Glebov1997} with the same overlay as in (e,f). Two unit cells related by mirror symmetry are indicated by colored rhombs for each model, (g) $\sqrt{37}$, (h) $\sqrt{39}$. \label{fig:HAS}}
\end{figure}

The diffraction spots in the HAS diffraction patterns of the $\sqrt{37}$ and the $\sqrt{39}$ \cite{Glebov1997} show strongly varying brightness as reproduced in Fig.~\ref{fig:HAS} (e,f). Without a complete analysis of diffracted intensities, we discuss in the following whether these brightness differences corroborated the structural motives for the wetting layer proposed either by us or by Glebov et al. \cite{Glebov1997}. The structural motives for the wetting layer proposed by Glebov et al. are reproduced in Fig.~\ref{fig:HAS} (c) and (d). We may associate with these molecular structures a 2D Bravais lattice, with each lattice point sitting in a center of an ice ring (or on every second molecule). We calculate the positions of the diffraction spots for this 2D-lattice from the simple rule \cite{Ertl/Kueppers}:

\begin{equation*}
\left(
\begin{array}{r}
    {\bf a}_1^*\\
    {\bf a}_2^*\\
\end{array}
\right) = \left(M^{-1}\right)^T \left(
\begin{array}{r}
    {\bf s}_1^*\\
    {\bf s}_2^*\\
\end{array}
\right).
\end{equation*}

Here $M$ is the matrix that describes the relation of the superstructure with respect to the ice lattice in real space, ${\bf a}_1^*$ and ${\bf a}_2^*$ are primitive translations of the ice lattice in reciprocal space (so any linear combination of these two describes a reflex in diffraction), and ${\bf s}_1^*$ and ${\bf s}_2^*$ are the primitive translations of the superstructure in reciprocal space where the linear combination create all spots depicted in Fig.~\ref{fig:HAS} (g), (h). The respective matrices according to the model proposed in \cite{Glebov1997}are given by

\begin{equation*}
M_{\sqrt{37},{\rm H}_2{\rm O}}^{\rm HAS} = \begin{pmatrix} 4 & 3\\  -3 & 1 \\ \end {pmatrix};  M_{\sqrt{39},{\rm H}_2{\rm O}}^{\rm HAs} = \begin{pmatrix}  4 & 4\\ -4 & 0\\ \end{pmatrix} 
\end{equation*}

The result of this calculation is visualized in Fig.~\ref{fig:HAS} (g) and (h) (blue symbols). Due to the existence of two rotational domains we obtain two unit cells in reciprocal space. The resulting locations in reciprocal space are circled in the experimental data in Fig.~\ref{fig:HAS} (e,f) (blue circles). One would expect these spots to be bright within the full superstructure diffraction pattern due to the coherent addition of contributions from molecules associated with these lattice points within the entire superstructure unit cell. In fact, they are only of low or mediocre intensity. As a remark of caution we note that in proceeding like this we neglect the additional distortions in the wetting layer which give rise to the superstructure as well as structure factor effects caused by the fact that two molecules are associated to each point of our two 2D-lattices or associated with the orientation of individual molecules.

Next we consider our structural models shown in Fig.~\ref{fig:HAS} (a) and (b). One may also associate a 2D Bravais lattice to the ridge phase of the $\sqrt{37}$ and the $\sqrt{39}$ by associating a lattice point to each center of an ice ring and extending this points into the space with the $\sqrt{3}$ motive. This lattice only partly represents the full wetting layer structure, but still most of the water molecules can be linked to this lattice. Here the respective matrices are

\begin{equation*}
M_{\sqrt{37},{\rm H}_2{\rm O}}^{\rm STM} = \begin{pmatrix} 4 & 2\\ -2 & 2\\ \end {pmatrix};  M_{\sqrt{39},{\rm H}_2{\rm O}}^{\rm STM} = \begin{pmatrix}  4 & 3\\ -3 & 1\\ \end{pmatrix}
\end{equation*} 

Again we calculate the diffraction spots for this 2D Bravais lattice [Fig.~\ref{fig:HAS} (g) and (h), red symbols] and circle them red in Fig.~\ref{fig:HAS} (e) and (f). The spots are among the brightest spots in the entire diffraction pattern. This fact lends additional credit to the structural motive proposed by us for the ridge phase.

\begin{figure}[htb]
\includegraphics[width=\textwidth]{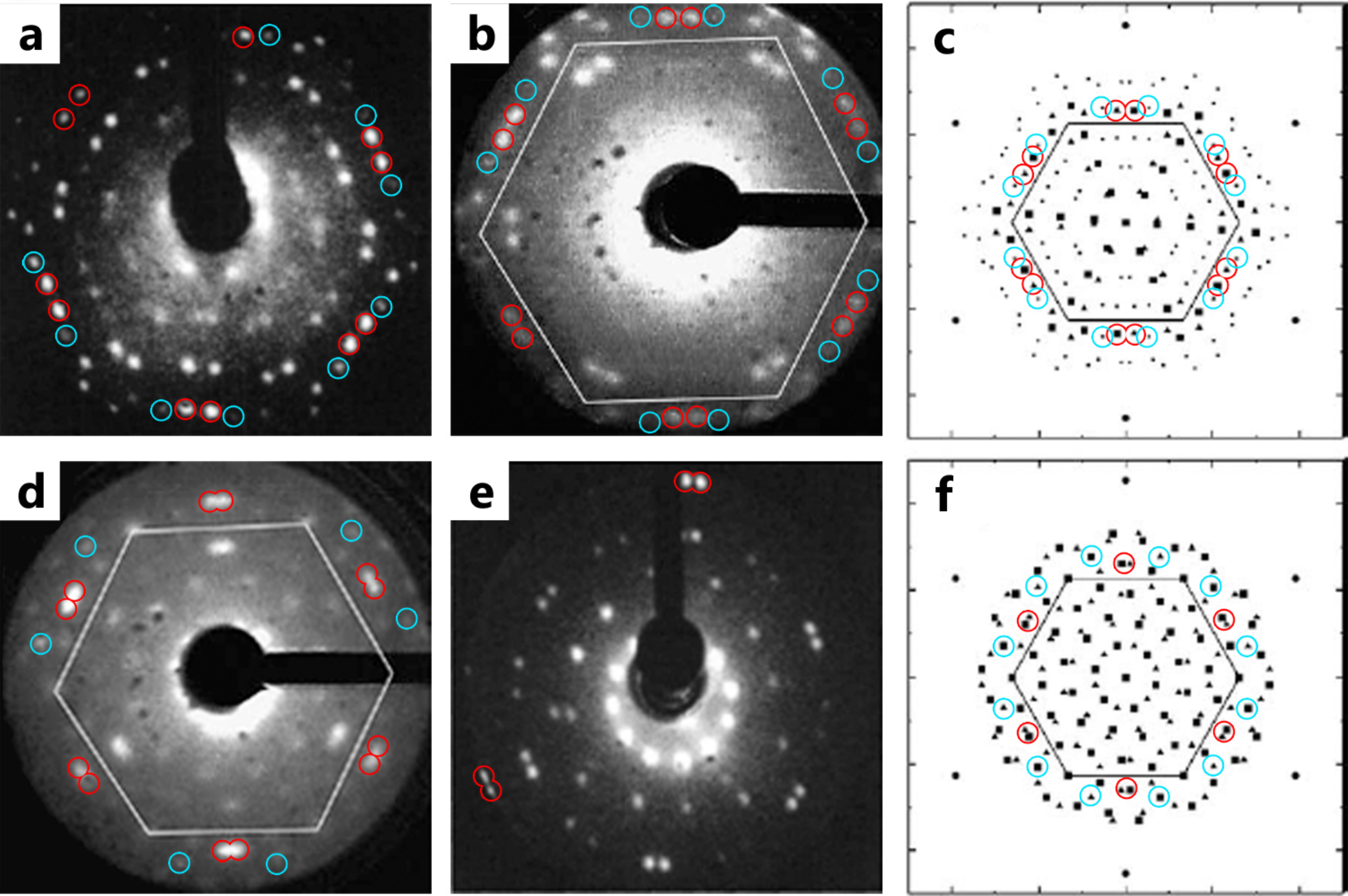}
\caption{LEED pattern and corresponding pattern generated from the reciprocal lattice of the two rotational domains (squares, circles) of (a-c) the $\sqrt{37}$ and (d-f) the $\sqrt{39}$ as originally published in \cite{Harnett2003} and \cite{Zimbitas2005}. The blue circles mark the calculated spots of the reciprocal lattice (two rotational domains) formed by the water molecules alone according to the model in \cite{Glebov1997} whereas red circles are expected for the structures proposed by us. Electron energies: 27\,eV (a), 23\,eV (b), 29\,eV (d), 15\,eV(e) \label{fig:LEED}.}
\end{figure}

As with the HAS data, one can analyze LEED patterns of the two phases in terms of molecular reflexes of the water layer alone. At this point one has to take into account that the spot intensity may vary upon voltage changes (\textit{IV} LEED). Due to the fact that we do not have access to an extended set of data beyond the published one, we can consider only single frames. Nevertheless, the compliance described for HAS data remains valid for all LEED pictures shown in Fig.~\ref{fig:LEED}. The molecular reflexes of the partly populated ice lattice forming the ridge structure (red circles) appear significantly brighter than the estimated reflexes of the Glebov-model \cite{Glebov1997} (blue circles), both for the $\sqrt{37}$ and the $\sqrt{39}$ superstructure and electron energies of 23\,eV, 27\,eV (for the $\sqrt{37}$ phase), 15\,eV and 29\,eV (for the $\sqrt{39}$ wetting layer).

